\documentclass[aps,prb,twocolumn,groupedaddress]{revtex4-1}
\usepackage[dvipdfmx]{graphicx}
\usepackage{amsmath,amssymb}
\usepackage{mathrsfs}
\usepackage{bm}
\usepackage{braket}
\newcommand{\Res}{\mathop{\rm Res}\limits}
\renewcommand{\Re}{{\rm Re}}
\newcommand{\eff}{{\rm eff}}

\begin{document}


\title{Stabilization of prethermal Floquet steady states in a periodically driven dissipative Bose-Hubbard model}
\author{Koudai Iwahori}
\email[]{iwahori@scphys.kyoto-u.ac.jp}
\author{Norio Kawakami}
\affiliation{Department of Physics, Kyoto University, Kyoto 606-8502, Japan}
\date{\today}

\begin{abstract}
We discuss the effect of dissipation on heating which occurs in periodically driven quantum many body systems.
We especially focus on a periodically driven Bose-Hubbard model coupled to an energy and particle reservoir.
Without dissipation, this model is known to undergo parametric instabilities which can be considered as an initial stage of heating.
By taking the weak on-site interaction limit as well as the weak system-reservoir coupling limit,
we find that parametric instabilities are suppressed if the dissipation is stronger than the on-site interaction strength and stable steady states appear.
Our results demonstrate that periodically-driven systems can emit energy, which is absorbed from external drivings, to the reservoir so that they can avoid heating.
\end{abstract}


\maketitle

\section{Introduction}
Long time behavior of periodically driven quantum systems has attracted particular attention due to the success of the Floquet engineering of matters\cite{Linger2007,Struck2012,Aidelsburger2013a,Aidelsburger2013b,Miyake2013,Aidelsburger2015,Wang2013,Ishikawa2014,Sie2015}.
Dynamics of such system is described by the Floquet effective Hamiltonian which is defined by $U(T)=\exp[-iH_{\eff}T]$, where $U(t)$ is the time evolution operator of the system and $T$ is the period of the external field.
This Floquet effective Hamiltonian is usually obtained by truncating the Magnus expansion and the truncated Floquet effective Hamiltonian can have non-trivial properties such as topologically non-trivial band structures \cite{Oka2009,Lindner2011,Kitagawa2011} which the original Hamiltonian does not have.

Recent studies have carefully investigated essential features of the Floquet effective Hamiltonian of quantum {\it isolated} non-integrable systems and revealed that the system absorbs energy from the external field and steady states realized after a long time are the trivial infinite temperature state\cite{DAlessio2013,Lazarides2014b,Ponte2015}.
Further studies have also shown that the energy absorption rate is small and there exists quasi-steady states (i.e. their life time is finite) which have finite temperature properties of the truncated Floquet effective Hamiltonian\cite{Abanin2015a,Abanin2015c,Mori2016,Kuwahara2016,Canovi2016,Lindner2016,Weidinger2016,Chandran2016}.
These facts suggest that there remains a possibility that the life time of quasi-steady states in {\it isolated} systems becomes infinite in {\it dissipative} systems because the system can emit the energy to the reservoir.
Solid state systems irradiated by intense laser lights are essentially a quantum dissipative system and in this sense, understanding the effect of dissipation on the heating in periodically driven quantum systems is a fundamentally important issue.

There has been also intensive research on periodically driven {\it dissipative} quantum systems\cite{Zerbe1995,Kohler1997,Breuer2000,Kohn2001,Hone2009,Ketzmerick2010,Iadecola2013,Langemeyer2014,Iadecola2015a,Iadecola2015b,Shirai2015a,Dehghani2014,Liu2015,Seetharam2015,Shirai2016,Iwahori2016}.
Recent studies have revealed that long time asymptotic states of periodically driven {\it dissipative} quantum systems are effectively described by the Gibbs distribution of the Floquet effective Hamiltonian in the high frequency regime of the external field\cite{Seetharam2015,Shirai2016,Iwahori2016}.
However, these studies are restricted to integrable systems or numerical calculations for a small non-integrable spin chain.
Moreover, most of previous studies concentrated on the case where the system reservoir coupling is taken to be infinitesimal compared with any other energy scales of the system,
therefore it is not clear how the heating which is caused by periodic drivings is affected by the dissipation.

Motivated by these backgrounds, we investigate the effect of dissipation on the heating in the periodically driven quantum systems.
We consider a periodically driven Bose-Hubbard model coupled to a reservoir and the on-site interaction is treated by the Hartree-Fock-Bogoliubov approximation. If there is no dissipation, this system is known to exhibit parametric instabilities with an infinitesimal on-site interaction\cite{Creffield2009,Bukov2015,Lellouch2016} 
and recent studies have revealed that the heating before the quasi steady state is governed by the parametric instabilities\cite{Chandran2016,Weidinger2016}.
By taking the weak on-site interaction limit as well as the weak system-reservoir coupling limit, we find that the Floquet steady state is stabilized if the dissipation is stronger than the on-site interaction strength.
Our results demonstrate that heating which is inevitable in the periodically driven {\it isolated} quantum systems can be suppressed by introducing dissipation and we can obtain the low energy properties of the Floquet effective Hamiltonian in the prethermal Floquet steady state if the dissipation is strong enough to overcome the heating rate.

In the next section, we show the model considered in this paper and derive the equation of motion of the system within the Hartree-Fock-Bogoliubov approximation.
In Sec. \ref{Sec_WeakCouplingLimit}, the equation of motion in the weak system-reservoir coupling and the weak on-site interaction limit is described,
then we discuss the stability of the system and properties of the steady state in Sec. \ref{Sec_Stability}.
A summary of our results is presented in Sec. \ref{Sec_Summary}.

\section{Setup}\label{Sec_Setup}
We study a periodically driven Bose-Hubbard model on a $d$-dimensional cubic lattice which couples to a reservoir.
The Hamiltonian of the total system is given by $H_{\text{tot}}(t)=H_{S}(t)+H_{R}+H_{I}$, where $H_{R}$ and $H_{S}(t)$ are the Hamiltonian of the reservoir and the system, respectively, and $H_{I}$ describes the interaction between them.
The Hamiltonian of the system is the Bose-Hubbard model driven by a time periodic force $\bm{E}(t)$:
\begin{equation}
  \begin{aligned}
    H_{S}(t) =& -t_{0}\sum_{\braket{i,j}}B^{\dagger}_{i}B_{j}+\frac{U}{2}\sum_{j}n_{j}(n_{j}-1)\\
	          &+ \sum_{j}(\bm{E}(t)\cdot \bm{r}_{j}n_{j} -\mu n_{j}),
  \end{aligned}
\end{equation}
where $B_{j}$ is the boson annihilation operator for particle on site $\bm{r}_{j}$, $n_{j}=B^{\dagger}_{j}B_{j}$ is the number operator, and $\braket{i,j}$ represents a pair of nearest neighbor sites.
The number of lattice site is taken as $N^{d}$ and $\mu$ is the chemical potential.
The external force has a period of $2\pi /\Omega$ and the time average of the external force is zero : $\bm{E}(t+2\pi /\Omega) = \bm{E}(t),\ \int_{0}^{2\pi /\Omega}dt\bm{E}(t)=\bm{0}$.
The Hamiltonian of reservoir and the interaction between the system and the reservoir are taken as
\begin{equation}
    \begin{aligned}
        &H_{R}=\sum_{k}\omega_{k}A^{\dagger}_{k}A_{k}\\
        &H_{I}=\sum_{k,j}\bigg[ \frac{\lambda_{k,j}}{\sqrt{L}}A^{\dagger}_{k}B_{j}+H.c. \bigg],
    \end{aligned}
\end{equation}
where $A_{k}$ describes the boson in the reservoir and $\lambda_{k,j}$ determines the coupling between the system and reservoir.
We consider the continuum limit $\sum_{k}\frac{2\pi}{L}F(\omega_{k})=\int_{-\infty}^{\infty}d\omega D(\omega)F(\omega)$, where $D(\epsilon)$ is the density of states of the reservoir and $F(\epsilon)$ is an arbitrary function.
The system and the reservoir exchange particles and energy because $H_{I}$ is a bilinear form of the Bose operators.

We consider the BEC state and separate out the condensate part as 
\begin{equation}
    \begin{aligned}
        &A_{k}=\Psi_{k}+a_{k}\\
        &B_{j}=\Phi_{j}+b_{j},
    \end{aligned}
\end{equation}
where $\Psi_{k} = \braket{A_{k}}$ and $\Phi_{j} = \braket{B_{j}}$ describe macroscopic Bose fields.
The initial state of the reservoir is taken as the equilibrium state with the temperature $T$, that is, $\braket{a^{\dagger}_{k}a_{k}}=f_{T}(\omega_{k})$, where $f_{T}(\omega)=1/(e^{\omega/T}-1)$ is the Bose distribution function (setting $k_{B}=1$ and $\hbar=1$).
We take the chemical potential of the reservoir as zero and thereby the energy of the reservoir is positive $\omega_{k} \geq 0$.
The initial state of the macroscopic Bose field in the reservoir is taken as $\Psi_{k}=\Psi\delta_{k,k_{0}}$, where $k_{0}$ specifies the smallest energy mode of the reservoir ($\text{min}_{k}\omega_{k}=\omega_{k_{0}}$ ).

The Heisenberg equation of the total system is written as
\begin{align}
  &i\frac{d}{dt}A_{k}(t)=\omega_{k}A_{k}(t)+\sum_{j}\frac{\lambda_{k,j}}{\sqrt{L}}B_{j}(t) \label{Eq_ReservoirEOM} \\
  &\begin{aligned}
    i\frac{d}{dt}B_{j}(t) =&-t_{0}\sum_{\braket{i,j}}B_{i}(t)+(\bm{E}(t)\cdot\bm{r}_{j}-\mu)B_{j}(t)\\
                           & +UB^{\dagger}_{j}(t)B_{j}(t)B_{j}(t)+\sum_{k}\frac{\lambda_{k,j}^{\ast}}{\sqrt{L}}A_{k}(t).\label{Eq_SystemEOM}
  \end{aligned}
\end{align}
We treat the on-site interaction term by the Hartree-Fock-Bogoliubov approximation\cite{Griffin1996}:
\begin{equation}
  \begin{aligned}
    B^{\dagger}_{j}B_{j}B_{j} \simeq & (|\Phi_{j}|^{2}+2\braket{b^{\dagger}_{j}b_{j}})\Phi_{j}+\braket{b_{j}b_{j}}\Phi^{\ast}_{j}\\
	                                 &+2(|\Phi_{j}|^{2}+\braket{b^{\dagger}_{j}b_{j}})b_{j}+(\Phi_{j}^{2}+\braket{b_{j}b_{j}})b^{\dagger}_{j}.
  \end{aligned}
\end{equation}
The Hartree-Fock-Bogoliubov approximation cannot describe the scattering of quasi-particles correctly.
However, within the weak on-site interaction limit which we will consider later, this process is irrelevant\cite{Bukov2015,Lellouch2016}.
Eq. (\ref{Eq_ReservoirEOM}) is integrated as
\begin{equation}
    A_{k}(t) = A_{k}(0) e^{-i\omega_{k}t}-i\sum_{j}\int_{0}^{t}dt^{\prime}\frac{\lambda_{k,j}}{\sqrt{L}}e^{-i\omega_{k}t^{\prime}}B_{j}(t-t^{\prime}). \label{Eq_SolReservoir}
\end{equation}
Substituting Eq. (\ref{Eq_SolReservoir}) into Eq. (\ref{Eq_SystemEOM}), we can obtain the closed form of the equation of motion for the system.
We here also apply a gauge transformation $B_{j}(t) \rightarrow B_{j}(t)\exp[-i\bm{\mathcal{A}}(t) \cdot \bm{r}_{j}]$, where $\bm{\mathcal{A}}(t)=\int_{0}^{t}dt^{\prime}\bm{E}(t^{\prime})$.
Then, we have the following equation of motion:
\begin{align}
  &\begin{aligned}
    i\frac{d}{dt}\Phi_{j}(t) =& -t_{0}\sum_{\braket{i,j}}e^{i\bm{\mathcal{A}}(t)\cdot(\bm{r}_{j}-\bm{r}_{i})}\Phi_{i}(t)-\mu \Phi_{j}(t)\\
                              &+U \big[ (n_{j}(t)+2\tilde{n}_{j}(t))\Phi_{j}(t)+\tilde{m}_{j}(t)\Phi^{\ast}_{j}(t) \big]\\
							  &-i\sum_{j^{\prime}}\int_{0}^{t}dt^{\prime}\Gamma_{jj^{\prime}}(t,t^{\prime})\Phi_{j^{\prime}}(t-t^{\prime})+i\Xi_{j}(t)\label{Eq_ClassicalEOM}
  \end{aligned}\\
  &\begin{aligned}
      i\frac{d}{dt}b_{j}(t)= & -t_{0}\sum_{\braket{i,j}}e^{i\bm{\mathcal{A}}(t)\cdot(\bm{r}_{j}-\bm{r}_{i})}b_{i}(t)-\mu b_{j}(t)\\
	                         &+2U(n_{j}(t)+\tilde{n}_{j}(t))b_{j}(t)\\
							 &+U(m_{j}(t)+\tilde{m}_{j}(t))b^{\dagger}_{j}(t)\\
							 &-i\sum_{j^{\prime}}\int_{0}^{t}dt^{\prime}\Gamma_{jj^{\prime}}(t,t^{\prime})b_{j^{\prime}}(t-t^{\prime})+i\xi_{j}(t),\label{Eq_QuantumEOM}
  \end{aligned}
\end{align}
where $\Gamma_{jj^{\prime}}(t,t^{\prime})=\sum_{k}\lambda^{\ast}_{k,j}(t)\lambda_{k,j^{\prime}}(t-t^{\prime})e^{-i\omega_{k}t}/L$ and $\Xi_{j}(t) = -i\sum_{k}\lambda_{k,j}^{\ast}(t)e^{-i\omega_{k}t}\Psi_{k}(0)/\sqrt{L},\ \xi_{j}(t)=-i\sum_{k}\lambda_{k,j}^{\ast}(t)e^{-i\omega_{k}t}a_{k}(0)/\sqrt{L}$.
$\lambda_{k,j}(t)$ is defined by $\lambda_{k,j}(t) = \lambda_{k,j}\exp[-i\bm{\mathcal{A}}(t)\cdot\bm{r}_{j}]$.
We have also introduced local densities:
\begin{equation}
  \begin{aligned}
    n_{j}(t)&=|\Phi_{j}(t)|^{2}\\
	m_{j}(t)&=\Phi_{j}(t)^{2}\\
	\tilde{n}_{j}(t)&=\braket{b^{\dagger}_{j}(t)b_{j}(t)}\\
	\tilde{m}_{j}(t)&=\braket{b_{j}(t)b_{j}(t)}.
  \end{aligned}
\end{equation}
The last line in Eq. (\ref{Eq_ClassicalEOM}) and Eq. (\ref{Eq_QuantumEOM}) describes a dissipation to the reservoir and a noise from the reservoir.

\section{Equation of motion in the weak coupling limit}\label{Sec_WeakCouplingLimit}
We solve Eqs. (\ref{Eq_ClassicalEOM}) and (\ref{Eq_QuantumEOM}) in both weak on-site interaction and weak system-reservoir coupling limit.
We here write down the equation of motion in the interaction picture.
Without the system-reservoir coupling and the on-site interaction, the solutions of Eqs. (\ref{Eq_ClassicalEOM}) and (\ref{Eq_QuantumEOM}) are $\Phi_{\bm{p}}(t) = \exp[-iE_{\bm{p}}t-i\theta_{\bm{p}}(t)]\Phi_{\bm{p}}(0)$ and $b_{\bm{p}}(t)=\exp[-iE_{\bm{p}}t-i\theta_{\bm{p}}(t)]b_{\bm{p}}(0)$, where $\Phi_{\bm{p}}$ and $b_{\bm{p}}$ are the Bloch state:
\begin{equation}
  \begin{aligned}
    \Phi_{\bm{p}}(t)&=\sum_{j}\frac{e^{-i\bm{p}\cdot\bm{r}_{j}}}{\sqrt{N}}\Phi_{j}(t)\\
    b_{\bm{p}}(t)&=\sum_{j}\frac{e^{-i\bm{p}\cdot\bm{r}_{j}}}{\sqrt{N}}b_{j}(t).
  \end{aligned}
\end{equation}
The quasi-energy of the system $E_{\bm{p}}$ and the time dependent phase $\theta_{\bm{p}}(t)$ are
\begin{equation}
  \begin{aligned}
    E_{\bm{p}} &= -2t_{0}\sum_{\alpha=1}^{d}|\mathcal{J}_{\alpha}^{(0)}|\cos(p_{\alpha}-q_{\alpha})-\mu\\
	\theta_{\bm{p}}(t) &= -t_{0}\int_{0}^{t} \Big[ \sum_{\alpha=1}^{d}\big( e^{i(p_{\alpha}-\mathcal{A}_{\alpha}(t^{\prime}))} + c.c. \big) -E_{\bm{p}} \Big]dt^{\prime},
  \end{aligned}
\end{equation}
where $\mathcal{J}_{\alpha}^{(n)}$ is the Fourier coefficient of $\exp[-i\mathcal{A}_{\alpha}(t)]$ and $\mathcal{J}_{\alpha}^{(0)}=|\mathcal{J}_{\alpha}^{(0)}|\exp[-iq_{\alpha}]$.
If the time reversal symmetry of the Floquet effective Hamiltonian is broken, we have $q_{\alpha} \neq 0,\pi$\cite{Sacha2012,Struck2012}.
$b_{\bm{p}}(t)$ is an annihilation operator of a Floquet state.

We define the interaction picture in a usual way \cite{Bilitewski2015}:
\begin{equation}
  \begin{aligned}
    \Phi_{\bm{p}}^{(I)}(t) = e^{iE_{\bm{p}}t+i\theta_{\bm{p}}(t)}\Phi_{\bm{p}}(t)\\
    b_{\bm{p}}^{(I)}(t) = e^{iE_{\bm{p}}t+i\theta_{\bm{p}}(t)}b_{\bm{p}}(t).\label{Eq_InteractionPicture}
  \end{aligned}
\end{equation}
We take the initial state of the macroscopic Bose field as $\Phi_{\bm{p}}(0)=\Phi\delta_{\bm{p},\bm{q}}$, where $\bm{q}$ labels the smallest energy Bloch state of the system (i.e. $\text{min}_{\bm{p}}E_{\bm{p}}=E_{\bm{q}}$).
We also assume that the initial state of the system does not have anomalous average, that is, $\braket{b_{\bm{p}}(0)b_{\bm{k}}(0)}=0$.
In the following, the frequency of the external force is taken in the range of $\Delta<\Omega<2\Delta$, where $\Delta = \text{max}_{\bm{p},\bm{k}}|E_{\bm{p}}-E_{\bm{k}}|$ is the single particle band width of the system.
Therefore, there is no resonance in the single particle level.
We define sets $\mathcal{S}(\omega):=\{ \bm{p}\in\mathcal{K}^{d} | E_{\bm{p}}-E_{\bm{q}} = \omega \}$ and $\mathcal{R}= \{ \bm{p}\in\mathcal{K}^{d} | E_{\bm{p}}-E_{\bm{q}}=\Omega /2 +O(U) \}$, where $\mathcal{K}^{d}$ is the discrete $d$-dimensional quasi-momentum space.
$\mathcal{S}(\omega)$ describes a set of states whose quasi-energy is $\omega$.
In the weak system-reservoir coupling and the weak on-site interaction limit, $\mathcal{R}=\emptyset$ or there exists a $\omega=E_{\bm{p}^{\ast}} \ (\bm{p}^{\ast}\in\mathcal{K}^{d})$ that satisfies $\mathcal{R}=\mathcal{S}(\omega)\neq\emptyset$.
The parametric instability occurs in $\bm{p}\in\mathcal{R}$.

Let us now consider the weak system-reservoir coupling limit (van Hove limit\cite{VanHove1955}) as done in many of previous papers\cite{Shirai2015a,Liu2015,Iadecola2015a,Shirai2016,Iwahori2016}.
Here, we also take the weak on-site interaction limit. This treatment enables us to discuss the effect of dissipation on the heating caused by the on-site interaction.
The meaning of the weak system-reservoir coupling limit and the weak on-site interaction limit is as follows:
We define the dissipation rate $\gamma = \max_{\bm{p}} \gamma_{\bm{p}}$ and the energy spacing $\delta=\text{min}_{E_{\bm{p}}\neq E_{\bm{k}}}|E_{\bm{p}}-E_{\bm{k}}|$, where $\gamma_{\bm{p}}=\int_{0}^{\infty}dt\sum_{n,m,\bm{p}^{\prime}}|\Gamma^{nm}_{\bm{p}\bm{p}^{\prime}}(t)|$ is the dissipation rate of the Bloch state $\bm{p}$ and $\Gamma^{nm}_{\bm{p}\bm{p}^{\prime}}(t)$ is defined in Eq (\ref{Eq_AppDefs}) in Appendix.
Then, take $\lim_{\gamma t,\ Ut \rightarrow \infty}\lim_{\gamma/\delta,\ U/\delta \rightarrow 0}$.
In this limit, we obtain the following equation of motions in the interaction picture (for details of the derivation, see Appendix):
\begin{widetext}
  \begin{equation}
    \begin{aligned}
      &i\frac{d}{dt}\Phi^{(I)}_{\bm{q}}(t)=u(t)\Phi^{(I)}_{\bm{q}}(t)+v(t)\Phi^{(I)\ast}_{\bm{q}}(t)
                                         -i\int_{0}^{t}dt^{\prime}\Gamma_{\bm{q}\bm{q}}(t^{\prime})\Phi^{(I)}_{\bm{q}}(t-t^{\prime})+i\Xi^{(I)}_{\bm{q}}(t)\\
      &i\frac{d}{dt}b^{(I)}_{\bm{p}}(t)=\tilde{u}(t)b^{(I)}_{\bm{p}}(t)+\tilde{v}_{\bm{p}}(t)b^{(I)\dagger}_{2\bm{q}-\bm{p}}(t)
                                      -i\sum_{\bm{p}^{\prime}\in\mathcal{S}(E_{\bm{p}})}\int_{0}^{t}dt^{\prime}\Gamma_{\bm{p}\bm{p}^{\prime}}(t^{\prime})b^{(I)}_{\bm{p}^{\prime}}(t-t^{\prime})+i\xi^{(I)}_{\bm{p}}(t),
    \end{aligned}\label{Eq_WeakCouplingLimitEOM}
  \end{equation}
  where $u(t)=U(n_{0}(t)+\tilde{n}(t))$ and $\tilde{u}(t)=2U(n_{0}(t)+\tilde{n}(t))$,
  \begin{equation}
      \begin{aligned}
        &v(t) = U \big[ \tilde{m}_{\bm{q}}(t) +\sum_{\bm{p}\in\mathcal{R}}(h_{\bm{p},\bm{q}}^{(1)})^{\ast}\tilde{m}_{\bm{p}}(t)e^{-2i(E_{\bm{p}}-E_{\bm{q}}-\Omega /2)t} \big]\\
        \tilde{v}_{\bm{p}}(t) &= \begin{cases}
                                   U \big[ m_{0}(t) + \tilde{m}_{\bm{q}}(t) +\sum_{\bm{k}\in\mathcal{R}}(h^{(1)}_{\bm{k},\bm{q}})^{\ast}(t)e^{-2i(E_{\bm{p}}-E_{\bm{q}}-\Omega /2)t} \tilde{m}_{\bm{k}}(t) \big] & \text{for } \bm{p}=\bm{q}\\[10pt]
                                   U \big[ (m_{0}(t)+\tilde{m}_{\bm{q}}(t)) h^{(1)}_{\bm{p},\bm{q}}e^{2i(E_{\bm{p}}-E_{\bm{q}}-\Omega /2)t}+\sum_{\bm{k}\in\mathcal{R}}h^{(0)}_{\bm{p},\bm{k}}\tilde{m}_{\bm{k}}(t) \big] & \text{for } \bm{p}\in\mathcal{R}\\[10pt]
                                   0 & \text{the others.}
                                 \end{cases}
    \end{aligned}\label{Eq_Coefficients}
  \end{equation}
\end{widetext}
We have here defined the mean fields:
\begin{equation}
  \begin{aligned}
    n_{0}(t)&=|\Phi_{\bm{q}}(t)|^{2} / N^{d}\\
    m_{0}(t)&=\big( \Phi_{\bm{q}}(t) \big) ^{2} / N^{d}\\
    \tilde{n}(t)&=\sum_{\bm{p}}\braket{b^{\dagger}_{\bm{p}}(t)b_{\bm{p}}(t)} / N^{d}\\
    \tilde{m}_{\bm{p}}(t)&= \braket{b^{(I)}_{\bm{p}}(t)b^{(I)}_{2\bm{q}-\bm{p}}(t)} / N^{d}.
  \end{aligned}\label{Eq_MeanField}
\end{equation}
$h^{(n)}_{\bm{p},\bm{k}}$ is the Fourier coefficient of $\exp[i(\theta_{\bm{p}}(t)+\theta_{2\bm{q}-\bm{p}}(t)-\theta_{\bm{k}}(t)-\theta_{2\bm{q}-\bm{k}}(t))]$.
The definition of all the coefficients is summarized in Appendix.
The macroscopic Bose fields whose quasi-momentum does not equal to $\bm{q}$ are zero at any time.
The first term in Eq. (\ref{Eq_WeakCouplingLimitEOM}) describes the normal mean field and only results in the energy shift.
The coefficients in the second term $v(t)$ and $\tilde{v}_{\bm{p}}(t)$ are written by using the square of the macroscopic Bose field (not the absolute value) and the anomalous average of the Bose operators.
This term describes a process of the pair creation or annihilation of the Bogoliubov phonon.
Note that there are two processes of the pair creation (annihilation) of Bogoliubov phonon.
One is the ordinary pair creation (annihilation) which the photon does not affect, the other is the pair creation (annihilation) in which a photon is absorbed (emitted).
The latter process is a cause of the parametric instability in this system \cite{Bukov2015,Lellouch2016}.
When the time reversal symmetry of the Floquet effective Hamiltonian is not broken, the photon mediated pair creation and annihilation process does not occur in the weak on-site interaction limit\cite{Lellouch2016} (i.e. $h^{(1)}_{\bm{p},\bm{k}}=0$ in our expression).
We can break the time reversal symmetry of the Floquet effective Hamiltonian by applying a multichromatic force\cite{Sacha2012,Struck2012}, and therefore we consider the external force which breaks the time reversal symmetry of the Floquet effective Hamiltonian.

\section{Stability of the Floquet state}\label{Sec_Stability}
We assume that the system is in the steady state of which the amplitude of the macroscopic Bose field $n_{0}(t)=|\Phi_{\bm{q}}|^{2}/N^{d}$ and the number of the excitations $\tilde{n}=\sum_{\bm{p}}\braket{b^{\dagger}_{\bm{p}}(t)b_{\bm{p}}(t)}/N^{d}$ are time independent.
We take the energy of the macroscopic Bose field in the reservoir as $\omega_{k_{0}} = E_{\bm{q}}+(n_{0}+2\tilde{n})U$ and the frequency of the external force as $\Omega=2(E_{\bm{p}}-E_{\bm{q}}+n_{0}U) \ (\bm{p}\in\mathcal{R})$. If there is a detuning, we can eliminate the detuning by taking a gauge transformation of $\Phi_{\bm{q}}$ and $b_{\bm{p}}$ respectively.

We apply the following gauge transformation:
\begin{equation}
  \begin{aligned}
    \Phi^{(I)}_{\bm{q}}(t) &\rightarrow \Phi^{(I)}_{\bm{q}}(t) e^{-i(n_{0}+2\tilde{n})Ut}\\
    b^{(I)}_{\bm{q}}(t) &\rightarrow b^{(I)}_{\bm{q}}(t) e^{-i(n_{0}+2\tilde{n})Ut}\\
    b^{(I)}_{\bm{p}\neq\bm{q}}(t) &\rightarrow b^{(I)}_{\bm{p}\neq\bm{q}}(t) e^{-2i(n_{0}+\tilde{n})Ut}.
  \end{aligned}\label{Eq_GaugeTransformation}
\end{equation}
Then, we have the equation of motion as
\begin{equation}
  \begin{aligned}
    &\begin{aligned}
	  i\frac{d}{dt}\Phi^{(I)}_{\bm{q}}(t)&= g\Phi^{(I)\ast}_{\bm{q}}(t)\\
                                        &-i\int_{0}^{t}dt^{\prime}\Gamma_{\bm{q}\bm{q}}(t^{\prime})\Phi^{(I)}_{\bm{q}}(t-t^{\prime})+i\Xi^{(I)}_{\bm{q}}(t)
    \end{aligned}\\
    &\begin{aligned}
      i\frac{d}{dt}b^{(I)}_{\bm{q}}(t)&=n_{0}Ub^{(I)}_{\bm{q}}(t)+ \big[ m_{0}U+g \big] b^{(I)\dagger}_{\bm{q}}(t)\\
                                     &-i\int_{0}^{t}dt^{\prime}\Gamma_{\bm{q}\bm{q}}(t^{\prime})b^{(I)}_{\bm{q}}(t-t^{\prime})+i\xi^{(I)}_{\bm{q}}(t)
    \end{aligned}\\
    &\begin{aligned}
      i\frac{d}{dt}b^{(I)}_{\bm{p}\neq\bm{q}}(t)&=\tilde{g}_{\bm{p}} b^{(I)\dagger}_{2\bm{q}-\bm{p}}(t)+i\xi^{(I)}_{\bm{p}}(t)\\
                                      &-i\sum_{\bm{p}^{\prime}\in\mathcal{S}(E_{\bm{p}})}\int_{0}^{t}dt^{\prime}\Gamma_{\bm{p}\bm{p}^{\prime}}(t^{\prime})b^{(I)}_{\bm{p}^{\prime}}(t-t^{\prime}).
    \end{aligned}
  \end{aligned}\label{Eq_FinalExpressionEOM}
\end{equation}
We have also transformed $m_{0}$ and $\tilde{m}_{\bm{p}}$, $\Gamma_{\bm{p}\bm{k}}(t)$, $\Xi^{(I)}_{\bm{q}}(t)$, $\xi^{(I)}_{\bm{p}}(t)$ in the same way as Eq. (\ref{Eq_GaugeTransformation}).
$g$ and $\tilde{g}_{\bm{p}}$ are defined as $g=U [\tilde{m}_{\bm{q}}+\sum_{\bm{p}\in\mathcal{R}}(h^{(1)}_{\bm{p},\bm{q}})^{\ast}\tilde{m}_{\bm{p}}]$ and $\tilde{g}_{\bm{p}\in\mathcal{R}}=U[(m_{0}+\tilde{m}_{\bm{q}})h^{(1)}_{\bm{p},\bm{q}}+\sum_{\bm{k}\in\mathcal{R}}h^{(0)}_{\bm{p},\bm{k}}\tilde{m}_{\bm{k}}]$, $\tilde{g}_{\bm{p}\not\in{R}}=0$.
We here also assume that the anomalous average $\tilde{m}_{\bm{p}}$ and the macroscopic Bose field $m_{0}$ are time independent.
We will see that this assumption is also satisfied if $n_{0}$ and $\tilde{n}$ are time independent.
Then, we can solve Eqs. (\ref{Eq_FinalExpressionEOM}) by the Laplace transformation.
Here we solve the equation of motion for $b^{(I)}_{\bm{p}\in\mathcal{R}}$.
The other equation of motion can be solved in the same way.
By applying the Laplace transformation, the equation of motion is transformed as
\begin{equation}
  \sum_{\bm{p}^{\prime}\in\mathcal{R}} \big( \delta_{\bm{p},\bm{p}^{\prime}}z+\Gamma_{\bm{p}\bm{p}^{\prime}}(z) \big)b^{(I)}_{\bm{p}^{\prime}}(z) +i\tilde{g}_{\bm{p}}b^{(I)\dagger}_{2\bm{q}-\bm{p}}(z)=b_{\bm{p}}+\xi^{(I)}_{\bm{p}}(z).\label{Eq_LaplaceTransformedEOM}
\end{equation}
Functions in the frequency space are specified by their argument $z$ while those in the time space by $t$.
We write the number of elements in $\mathcal{R}$ as $n$ and write each element as $\bm{p}_{i}\in\mathcal{R} \ (i=1,2,\cdots ,n)$.
We define $\bm{b}^{(I)}(z)=(b^{(I)}_{\bm{p}_{1}}(z), \cdots , b^{(I)}_{\bm{p}_{n}}(z), b^{(I)\dagger}_{\bm{p}_{1}}(z), \cdots ,b^{(I)\dagger}_{\bm{p}_{n}}(z))^{\text{T}}$ and
$\bm{b}=(b_{\bm{p}_{1}}, \cdots , b_{\bm{p}_{n}}, b^{\dagger}_{\bm{p}_{1}}, \cdots ,b^{\dagger}_{\bm{p}_{n}})^{\text{T}}$,
$\bm{\xi}(z)=(\xi^{(I)}_{\bm{p}_{1}}(z), \cdots , \xi^{(I)}_{\bm{p}_{n}}(z), \xi^{(I)\dagger}_{\bm{p}_{1}}(z), \cdots ,\xi^{(I)\dagger}_{\bm{p}_{n}}(z))^{\text{T}}$.
Then, we rewrite Eq. (\ref{Eq_LaplaceTransformedEOM}) in the matrix form:
\begin{equation}
  M(z)\bm{b}^{(I)}(z)=\bm{b}+\bm{\xi}(z),\label{Eq_LaplaceTransformedEOMMatrixForm}
\end{equation}
where the matrix $M(z)$ is defined through Eq. (\ref{Eq_LaplaceTransformedEOM}).
In this expression, the solution is obtained in the following way:
\begin{equation}
  \begin{aligned}
    \bm{b}^{(I)}(t)&=\sum_{l} \Res_{z=z_{l}}\big[ \big( M^{-1}(z)(\bm{b}+\bm{\xi}(z) \big) e^{zt} \big]\\
                   &+ \sum_{k}\sum_{n\in\mathbb{Z}} \Res_{z=\pm i(\omega_{k}-\epsilon_{\bm{p}}^{(n)})} \big[ M^{-1}(z)\bm{\xi}(z)e^{zt} \big],
  \end{aligned}\label{Eq_SteadyStateSolution1}
\end{equation}
where $z_{l}$ is the solution of $\det M(z)=0$ and $\epsilon_{\bm{p}}^{(n)}=E_{\bm{p}}+2(n_{0}+\tilde{n})U+n\Omega$.
We here introduce the energy cutoff of the system-reservoir coupling $\omega_{c}$ \cite{Shirai2016, Iwahori2016} and also consider the limit of $\gamma /\omega_{c},\ U/\omega_{c}\rightarrow 0$.
When we solve $\det M(z)=0$, we can take $\Gamma_{\bm{p}\bm{p}^{\prime}}(z)$ as $\Gamma_{\bm{p}\bm{p}^{\prime}}(0)$ because the solution of $\det M(z)=0$ are $z=O(\gamma)+O(U)$ and we have taken the limit of $\gamma /\omega_{c},\ U/\omega_{c}\rightarrow 0$.
Therefore, $\det M(z)=0$ reduces to a $2n$th polynomial equation and there are $2n$ solutions.
The second term in Eq. (\ref{Eq_SteadyStateSolution1}) describes the steady state.
If the the first term decays (i.e. $\Re[z_{l}]<0,\ \forall l$), we obtain the steady state solution in the long time limit $z_{l}t\rightarrow \infty$ as
\begin{equation}
  \bm{b}^{(I)}(t)=\sum_{k}\sum_{n\in\mathbb{Z}} \Res_{z=\pm i(\omega_{k}-\epsilon_{\bm{p}}^{(n)})} \big[ M^{-1}(z)\bm{\xi}(z)e^{zt} \big].
\end{equation}
Without dissipation (i.e. $U / \gamma \rightarrow\infty$), we have $z_{l}=\pm |\tilde{g}|$, and thereby we cannot obtain the steady state due to the occurrence of the parametric instability\cite{Bukov2015,Lellouch2016}.
In this regime, a number of excitations (Bogoliubov phonons whose momenta are in $\mathcal{R}$) are created by the external force and the mean field analysis is not applicable in the long time steady state\cite{Bukov2015,Lellouch2016}.
On the other hand, we have $\Re[z_{l}]<0$ without on-site interaction (i.e. $U/\gamma\rightarrow 0$) and the steady state is written in the same way as our previous results \cite{Iwahori2016}.
Therefore, there is a point where the stability and instability changes in $U/\gamma \in\mathbb{R}_{>0}$ and the parametric instabilities can be suppressed by the dissipation.
We calculate this value in Sec. \ref{Sec_Example}.
We can prove the same thing for the other equation of motion.

In the steady state, the time evolution of the operator $b^{(I)}_{\bm{p}}(t)$ is written in the following way:
\begin{equation}
  \begin{aligned}
    b^{(I)}_{\bm{p}}(t)=&\sum_{k,n,l} \frac{\tilde{\lambda}_{k,l,\bm{p}}^{(n)}}{\sqrt{L}}\frac{\tilde{\gamma}_{l,\bm{p}}e^{-i(\omega_{k}-\epsilon_{\bm{p}}^{(n)})t}}{\omega_{k}-\epsilon_{\bm{p}}^{(n)}-iz_{l}} a_{k}(0)\\
                        &+\sum_{k,n,l} \frac{\tilde{\nu}_{k,l,\bm{p}}^{(n)}}{\sqrt{L}}\frac{\tilde{u}_{l,\bm{p}}e^{i(\omega_{k}-\epsilon_{\bm{p}}^{(n)})t}}{\omega_{k}-\epsilon_{\bm{p}}^{(n)}+iz_{l}} a^{\dagger}_{k}(0),
  \end{aligned}\label{Eq_SteadyStateSolution2}
\end{equation}
where each coefficient is determined by $M^{-1}(z)$ and $z_{l}$.
$b^{(I)}_{\bm{p}}(t)$ is written in terms of $a^{\dagger}_{k}$ as well as $a_{k}$.
This is because there is a coupling between $b^{(I)}_{\bm{p}}(t)$ and $b^{(I)\dagger}_{2\bm{q}-\bm{p}}(t)$.
This coupling is induced by the on-site interaction $U$, and thereby the second term is $O(U/\gamma)$.
By using Eq. (\ref{Eq_SteadyStateSolution2}), the occupation number of the Floquet state and the anomalous average can be obtained as
\begin{equation}
  \begin{aligned}
    &\braket{b^{\dagger}_{\bm{p}}(t)b_{\bm{p}}(t)}=\sum_{n=0}^{\infty}w^{(n)}_{\bm{p}}f_{T}(E_{\bm{p}}+n\Omega)+c_{\bm{p}}\\
    &\braket{b^{(I)}_{\bm{p}}(t)b^{(I)}_{2\bm{q}-\bm{p}}(t)}=\sum_{n=0}^{\infty}\tilde{w}^{(n)}_{\bm{p}}f_{T}(E_{\bm{p}}+n\Omega)+\tilde{c}_{\bm{p}},
  \end{aligned}\label{Eq_OccupationNumber}
\end{equation}
where $w^{(n)}_{\bm{p}} \geq 0$ and $\tilde{w}^{(n)}_{\bm{p}}\in\mathbb{C}$ determine the contribution from the $n$th Floquet sideband $E_{\bm{p}}+n\Omega$.
$c_{\bm{p}}\geq 0$ and $\tilde{c}_{\bm{p}}\in\mathbb{C}$ are independent of the temperature of the reservoir.
The behavior of the first term in Eq. (\ref{Eq_OccupationNumber}) is the same as many of the previous results\cite{Seetharam2015,Iadecola2015a,Iadecola2015b,Iwahori2016}.
The second term comes from the on-site interaction effect and describes small number of excitations created by the external force.
Even if we take the zero temperature limit, the second term remains finite while the first term goes to zero.
This means that the reservoir can prevent the system from heating but there remains small number of excitations in the system.
We will show in the next section, $c_{\bm{p}}$ increases as the on-site interaction becomes stronger.

The occupation number of the Floquet states $\bm{p}\not\in\mathcal{R}$ can be obtained in the same way:
\begin{equation}
    \braket{b^{\dagger}_{\bm{p}}(t)b_{\bm{p}}(t)}=\sum_{n=0}^{\infty}w^{(n)}_{\bm{p}}f_{T}(E_{\bm{p}}+n\Omega) \ (\bm{p}\not\in\mathcal{R}).\\
\end{equation}
By utilizing our previous results\cite{Iwahori2016}, when the energy cutoff of the system-reservoir coupling $\omega_{c}$ is smaller than the frequency of the external force $\Omega$, 
we have $w^{(0)}_{\bm{p}}\simeq 1$ and $w^{(n\neq 0)}_{\bm{p}} \ll 1$ for states $E_{\bm{p}}<\omega_{c}$. Therefore the occupation number of the Floquet state whose quasi-energy is smaller than the energy cutoff of the system-reservoir coupling is effectively written by the Bose distribution function:
\begin{equation}
    \braket{b^{\dagger}_{\bm{p}}(t)b_{\bm{p}}(t)}\simeq f_{T}(E_{\bm{p}}) \ (\bm{p}\not\in\mathcal{R} \text{ and } E_{\bm{p}}<\omega_{c}).\\
\end{equation}
The other time correlation functions have the same structure.
As we can see, there are no excitations in states $\bm{p}\not\in\mathcal{R}$.
This is because the origin of the excitation $c_{\bm{p}}$ is the coupling between $b_{\bm{p}}$ and $b^{\dagger}_{2\bm{q}-\bm{p}}$ via a photon assist (i.e. pair creation/annihilation of Bogoliubov phonon with one photon absorption/emission) in this case.
If we consider a stronger on-site interaction than this case, the scattering of the quasi-particles are induced.
This effect causes the diffusion of excitations to all the states and may enable us to define the effective temperature which is obtained in the previous papers\cite{Breuer2000,Ketzmerick2010,Shirai2016}.

From these observations, we conclude that we can prevent periodically driven quantum systems from heating by introducing dissipation which overwhelms the heating rate.
In the realized steady state, we obtain low energy properties of the Floquet effective Hamiltonian.
Recent studies on periodically driven quantum {\it isolated} systems have revealed that there are two phases of heating\cite{Chandran2016,Weidinger2016}:
The initial stage of heating is governed by the parametric instabilities.
Then, the system reaches a quasi-steady state whose energy absorption rate is small.
After the quasi-steady state, the system starts to absorb energy again and goes to the infinite temperature state.
This heating is governed by the scattering of the excitations which are created in the initial stage of heating.
Comparing these previous results of {\it isolated} systems, our results demonstrate that the initial stage of heating is prevented and quantum phases which are expected to be realized from the Floquet effective Hamiltonian will be stabilized in the {\it prethermal} Floquet steady state\cite{Bukov2015,Lellouch2016}.
Note that our results do not directly mean that we can avoid the heating to the infinite temperature state.
However, from the discussion above, our results suggest that dissipation will also suppress the second stage of heating which is caused by the diffusion of the excitations to all the states because a number of excitations are suppressed by the dissipation and the dissipation channels exist in  all the states.

\section{Example}\label{Sec_Example}
\begin{figure}[t]
	\centering
	\includegraphics[width=80mm]{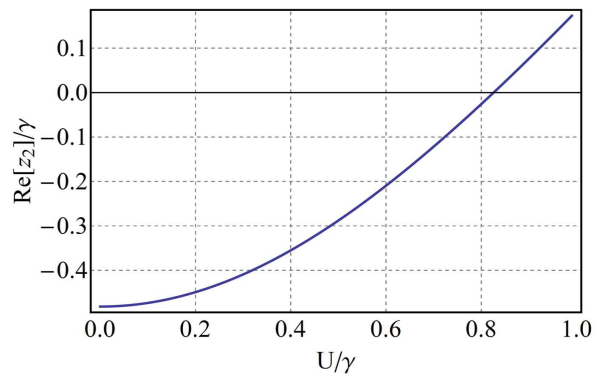}
	\caption{The plot of $\Re[z_{2}]$ as a function of $U/\gamma$.
             $\Re[z_{2}]$ gradually increases and crosses zero (i.e. the steady state becomes unstable) at $U/\gamma \simeq 0.82$.}
	\label{Fig_Stability}
\end{figure}

\begin{figure}[t]
	\centering
	\includegraphics[width=80mm]{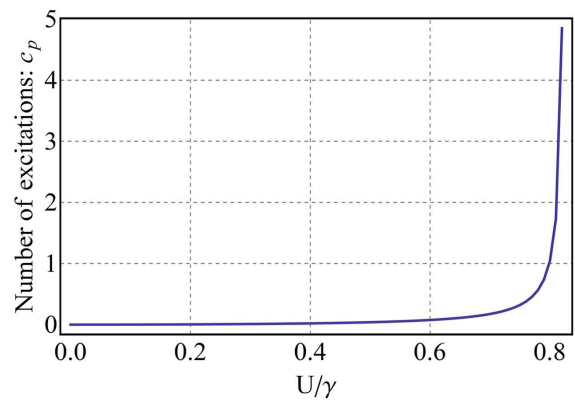}
	\caption{The number of excitations defined in Eq. (\ref{Eq_OccupationNumber}).
             As the strength of the on-site interaction increases, the number of excitations also increases.}
	\label{Fig_EffTemp}
\end{figure}

Here, we show an example of the above discussion.
We will discuss the stability of the steady state and calculate the number of excitations defined in Eq. (\ref{Eq_OccupationNumber}).
We consider a one-dimensional system and the external force is taken as $E(t)=E_{1}\cos(\Omega t)+E_{2}\cos(2\Omega t+\phi)$.
In this case, the Fourier coefficient of $\exp[-i\mathcal{A}(t)]$ is obtained as\cite{Sacha2012}
\begin{equation}
  \mathcal{J}^{(n)}=\sum_{m\in\mathbb{Z}}J_{n-2m}(\zeta_{1})J_{m}(\zeta_{2})e^{-im\phi},
\end{equation}
where $\zeta_{1}=E_{1}/\Omega$, $\zeta_{2}=E_{2}/\Omega$ and $J_{n}(\zeta)$ is the $n$th Bessel function.
The momentum of the BEC is obtained by $q=-\arg (\mathcal{J}^{(0)})$.
The system reservoir coupling which is defined in Sec. \ref{Sec_Setup} is taken as $\lambda_{k,j} = \lambda F(\omega_{k}) \delta_{j,0}$ and $D(\omega)F(\omega)^{2} = \theta(\omega)/(1+(\omega/\omega_{c})^{6})$, where $\theta(x)$ is the step function.
We define the dissipation rate by $\gamma = |\lambda|^{2}$.
For simplicity, we solve Eq. (\ref{Eq_FinalExpressionEOM}) within the Bogoliubov approximation.
We only consider the equation of motion for the state in $\mathcal{R}$ because the parametric instability does not occur in the other equation of motion.
Then, the equation of motion for the state $p\in\mathcal{R}$ is written as
\begin{equation}
  \begin{aligned}
    i\frac{d}{dt}b^{(I)}_{p}(t)=&m_{0}Uh^{(1)}_{p,q}b^{(I)}_{2q-p}(t)+i\xi^{(I)}_{p}(t)\\
                                &-i\sum_{p^{\prime}=p,2q-p}\int_{0}^{t}dt^{\prime}\Gamma_{pp^{\prime}}(t^{\prime})b^{(I)}_{p^{\prime}}(t-t^{\prime}).
  \end{aligned}
\end{equation}
The sum in the third term  $\sum_{p^{\prime}=p,2q-p}$ results from the degeneracy $E_{p}=E_{2q-p}$.
For simplicity, we neglect the sum for $p^{\prime}=2q-p$ and the equation of motion is
\begin{equation}
  i\frac{d}{dt}b^{(I)}_{p}(t)=ub^{(I)}_{2q-p}(t)-i\int_{0}^{t}dt^{\prime}\Gamma_{pp}(t^{\prime})b^{(I)}_{p}(t-t^{\prime})+i\xi^{(I)}_{p}(t),\label{Eq_ExampleEOM}
\end{equation}
where $u=m_{0}Uh^{(1)}_{p,q}$.
Formally, this treatment is justified by applying a perturbation which breaks degeneracy of $p,2q-p\in\mathcal{R}$ and keeps $E_{p}+E_{2q-p}-2E_{q}+2n_{0}U=\Omega$.
Then, after taking the weak system-reservoir coupling and the weak on-site interaction limit, we switch off the perturbation.
Eq. (\ref{Eq_ExampleEOM}) can be solved in the same way as Eq. (\ref{Eq_SteadyStateSolution2}) and (\ref{Eq_OccupationNumber}).
The solution of $\det M(z)=0$, which determines the stability of the system is explicitly written as
\begin{equation}
  \begin{aligned}
    z_{1,2}=-\frac{1}{2} \bigg[ & \Gamma_{p,p}+\Gamma_{2q-p,2q-p} \\
              &\pm \sqrt{(\Gamma_{p,p}-\Gamma_{2q-p,2q-p})^{2}+4|u|^{2}} \bigg],
  \end{aligned}
\end{equation}
where $\Gamma_{p,p}=\Gamma_{p,p}(z=0)$ and $\Re[\Gamma_{p,p}],\ \Re[\Gamma_{2q-p,2q-p}]>0$.
If $\Re[z_{1}]$ and $\Re[z_{2}]$ are both negative, the steady state is stable.
Note that $\Re[z_{1}]$ is always negative, therefore we discuss only $\Re[z_{2}]$.
The figure of $\Re[z_{2}]$ and $c_{\bm{p}}$ as a function of $U/\gamma$ is shown in Fig. \ref{Fig_Stability} and \ref{Fig_EffTemp}.
The parameters are taken as $(\zeta_{1}, \zeta_{2}, \phi, N, t_{0}/\Omega, \omega_{c}/\Omega, n_{0}) = (1.7965, 1, \pi /2, 101, 0.5213, 0.5, 1)$.
$\Re[z_2]$ gradually increases as $U/\gamma$ increases and at $U/\gamma \simeq 0.82$, $\Re[z_{2}]$ crosses zero and the steady state becomes unstable.
The number of excitations which are created by the external force also monotonically increases as the on-site interaction becomes strong.
At the stable/unstable transition point, the number of excitations $c_{\bm{p}}$ diverges.
Similar monotonic behaviors of $\Re[z_{2}]$ and $c_{\bm{p}}$ can be observed in other parameter sets, but the stable/unstable transition point is different depending on the parameters employed. 
The stable/unstable transition point is determined by the dissipation rate of states $p\in\mathcal{R}$ ($\Gamma_{p,p}$ and $\Gamma_{2q-p,2q-p}$). 
Therefore, the transition point especially depends on the details of the system reservoir coupling $\lambda_{k,j}$, which determines the dissipation rate of each Bloch state $p\in\mathcal{K}$. 


\section{Summary}\label{Sec_Summary}
To summarize, we have investigated steady states of a periodically driven dissipative Bose-Hubbard model especially focusing on whether the dissipation can prevent the system from  heating.
We have revealed that the parametric instability which is considered as an initial stage of the heating\cite{Chandran2016,Weidinger2016} can be prevented if the dissipation is stronger than the on-site interaction and stabilizes the steady states.
Our findings demonstrate that the low temperature properties of the Floquet effective Hamiltonian are obtained in the prethermal Floquet steady state if the dissipation overwhelms the heating rate. 
Our results also elucidate that the excitations which are created in the initial stage of heating are suppressed.
This fact suggests that the second stage of heating which is governed by the scattering of the excitations\cite{Weidinger2016} will be also suppressed by introducing dissipation, and thus the heating to the infinite temperature state will be avoided.
In order to check this suggestion, we have to go beyond the weak system-reservoir coupling and the weak on-site interaction limit.  This is left for  a future study.

\begin{acknowledgments}
This work was partly supported by a Grand-in-Aid for Scientific Research on Innovative Areas (JSPS KAKENHI Grant No. JP15H05855) and also JSPS KAKENHI (No. JP16K05501).
\end{acknowledgments}

\appendix
\begin{widetext}

\section{Derivation of eq. (\ref{Eq_WeakCouplingLimitEOM})}\label{appendix_a}
We derive the equation of motion in the weak on-site interaction and the weak system reservoir coupling limit.
In the interaction picture defined by Eq (\ref{Eq_InteractionPicture}), the equation of motion for the Bloch state is obtained in the following way:
\begin{equation}
    \begin{aligned}
        &\begin{aligned}
        i\frac{d}{dt}\Phi_{\bm{p}}^{(I)}(t) &= \frac{U}{N^{d}} \sum_{\bm{p}^{\prime},\bm{k},\bm{k}^{\prime}}\sum_{m\in\mathbb{Z}}\sum_{\bm{n}\in\mathbb{Z}^{d}}
                                                 \Big[ \big( n_{\bm{k},\bm{k}^{\prime}}+2\tilde{n}_{\bm{k},\bm{k}^{\prime}} \big) \Phi^{(I)}_{\bm{p}^{\prime}}(t)
                                                      + \tilde{m}_{\bm{p}^{\prime},\bm{k}^{\prime}} \Phi^{(I)\ast}_{\bm{k}}(t) \Big] \\
                                            & \qquad \qquad \qquad \qquad \qquad \times h^{(m)}_{\bm{p},\bm{p}^{\prime},\bm{k},\bm{k}^{\prime}} e^{i(E_{\bm{p}}+E_{\bm{k}}-E_{\bm{p}^{\prime}}-E_{\bm{k}^{\prime}}-m\Omega)t} \delta_{\bm{p}+\bm{k},\bm{p}^{\prime}+\bm{k}^{\prime}+2\pi\bm{n}}\\
                                            &-i\sum_{\bm{p}^{\prime}}\sum_{n,m\in\mathbb{Z}}e^{i(E_{\bm{p}}-E_{\bm{p}^{\prime}}+(n-m)\Omega)t}\int_{0}^{t}dt^{\prime}\Gamma_{\bm{p}\bm{p}^{\prime}}^{nm}(t^{\prime})e^{i(E_{\bm{p}^{\prime}}+m\Omega)t^{\prime}}\Phi^{(I)}_{\bm{p}^{\prime}}(t-t^{\prime})+i\Xi_{\bm{p}}^{(I)}(t)
        \end{aligned}\\
        &\begin{aligned}
        i\frac{d}{dt}b_{\bm{p}}^{(I)}(t) &= \frac{U}{N^{d}} \sum_{\bm{p}^{\prime},\bm{k},\bm{k}^{\prime}}\sum_{m\in\mathbb{Z}}\sum_{\bm{n}\in\mathbb{Z}^{d}}
                                                 \Big[ 2 \big( n_{\bm{k},\bm{k}^{\prime}}+\tilde{n}_{\bm{k},\bm{k}^{\prime}} \big) b^{(I)}_{\bm{p}^{\prime}}(t)
                                                      + \big( m_{\bm{p}^{\prime},\bm{k}^{\prime}}+\tilde{m}_{\bm{p}^{\prime},\bm{k}^{\prime}} \big) b^{(I)\dagger}_{\bm{k}}(t) \Big] \\
                                         &\qquad \qquad \qquad \qquad \qquad \times h^{(m)}_{\bm{p},\bm{p}^{\prime},\bm{k},\bm{k}^{\prime}} e^{i(E_{\bm{p}}+E_{\bm{k}}-E_{\bm{p}^{\prime}}-E_{\bm{k}^{\prime}}-m\Omega)t} \delta_{\bm{p}+\bm{k},\bm{p}^{\prime}+\bm{k}^{\prime}+2\pi\bm{n}} \delta_{\bm{p}+\bm{k},\bm{p}^{\prime}+\bm{k}^{\prime}+2\pi\bm{n}}\\
                                         &-i\sum_{\bm{p}^{\prime}}\sum_{n,m\in\mathbb{Z}}e^{i(E_{\bm{p}}-E_{\bm{p}^{\prime}}+(n-m)\Omega)t}\int_{0}^{t}dt^{\prime}\Gamma_{\bm{p}\bm{p}^{\prime}}^{nm}(t^{\prime})e^{i(E_{\bm{p}^{\prime}}+m\Omega)t^{\prime}}b_{\bm{p}^{\prime}}(t-t^{\prime})+i\xi_{\bm{p}}^{(I)}(t),
        \end{aligned}
    \end{aligned}
\end{equation}
where $n_{\bm{k},\bm{k}^{\prime}}$, $m_{\bm{k},\bm{k}^{\prime}}(t)$, $\tilde{n}_{\bm{k},\bm{k}^{\prime}}(t)$, $\tilde{m}_{\bm{k},\bm{k}^{\prime}}(t)$ are the mean fields in the interaction picture,
\begin{equation}
    \begin{aligned}
        &n_{\bm{k},\bm{k}^{\prime}}(t) = \Phi_{\bm{k}}^{\ast}(t)\Phi_{\bm{k}^{\prime}}(t),
            \quad m_{\bm{k},\bm{k}^{\prime}}(t) = \Phi^{(I)}_{\bm{k}}(t)\Phi^{(I)}_{\bm{k}^{\prime}}(t)\\
        &\tilde{n}_{\bm{k},\bm{k}^{\prime}}(t) = \braket{b^{\dagger}_{\bm{k}}(t)b_{\bm{k}^{\prime}}(t)},
            \quad \tilde{m}_{\bm{k},\bm{k}^{\prime}}(t) = \braket{b^{(I)}_{\bm{k}}(t)b^{(I)}_{\bm{k}^{\prime}}(t)}.
    \end{aligned}
\end{equation}
The dissipation $\Gamma^{nm}_{\bm{p}\bm{p}^{\prime}}(t)$ and the noise $\Xi^{(I)}_{\bm{p}}(t)$, $\xi^{(I)}_{\bm{p}}(t)$ are defined by
\begin{equation}
    \begin{aligned}
        &\Gamma^{nm}_{\bm{p}\bm{p}^{\prime}}(t) = \sum_{k}\frac{\lambda^{(m)\ast}_{k,\bm{p}}\lambda^{(n)}_{k,\bm{p}^{\prime}}}{L}e^{-i\omega_{k}t} \\
        &\Xi^{(I)}_{\bm{p}} = -i \sum_{n\in\mathbb{Z}}\frac{\lambda^{(n)\ast}_{k_{0},\bm{p}}}{\sqrt{L}}e^{-i(\omega_{k_{0}}-E_{\bm{p}}-n\Omega)t}\Psi \\
        &\xi^{(I)}_{\bm{p}} = -i \sum_{k}\sum_{n\in\mathbb{Z}}\frac{\lambda^{(n)\ast}_{k,\bm{p}}}{\sqrt{L}}e^{-i(\omega_{k}-E_{\bm{p}}-n\Omega)t}a_{k} \\
        &\lambda^{(n)}_{k,\bm{p}} = \int_{-\pi /\Omega}^{\pi / \Omega}dt \sum_{j}\frac{\lambda_{k,j}}{N^{d/2}}e^{i(\bm{p}-\bm{\mathcal{A}}(t))\cdot\bm{r}_{j}+i\theta_{\bm{p}}(t)+in\Omega t}.
    \end{aligned}\label{Eq_AppDefs}
\end{equation}
$h^{(m)}_{\bm{p},\bm{p}^{\prime},\bm{k},\bm{k}^{\prime}}$ is the Fourier coefficient of $\exp[i(\theta_{\bm{p}}(t)+\theta_{\bm{k}}(t)-\theta_{\bm{p}^{\prime}}(t)-\theta_{\bm{k}^{\prime}}(t))]$.
As we mentioned in the main text, the frequency of the external force is taken in the regime of $\Delta < \Omega < 2\Delta$, that is, there is no resonance in the single particle level.
In the weak system reservoir coupling and the weak on-site interaction limit, we can show that the only $n=m,\ E_{\bm{p}}=E_{\bm{p}^{\prime}}$ term in the dissipation term contribute to the dynamics.
We can also show similar things for the on-site interaction term.
Then, the equation of motion in the interaction picture reads
\begin{equation}
    \begin{aligned}
        &i\frac{d}{dt}\Phi^{(I)}_{\bm{q}}(t)=u(t)\Phi^{(I)}_{\bm{q}}(t)+v(t)\Phi^{(I)\ast}_{\bm{q}}(t)
                                         -i\int_{0}^{t}dt^{\prime}\Gamma_{\bm{q}\bm{q}}(t^{\prime})\Phi^{(I)}_{\bm{q}}(t-t^{\prime})+i\Xi^{(I)}_{\bm{q}}(t)\\
        &i\frac{d}{dt}b^{(I)}_{\bm{p}}(t)=\tilde{u}(t)b^{(I)}_{\bm{p}}(t)+\tilde{v}_{\bm{p}}(t)b^{(I)\dagger}_{2\bm{q}-\bm{p}}(t)
                                      -i\sum_{\bm{p}^{\prime}\in\mathcal{S}(E_{\bm{p}})}\int_{0}^{t}dt^{\prime}\Gamma_{\bm{p}\bm{p}^{\prime}}(t^{\prime})b^{(I)}_{\bm{p}^{\prime}}(t-t^{\prime})+i\xi^{(I)}_{\bm{p}}(t).
    \end{aligned}
\end{equation}
$u(t)$, $v(t)$, $\tilde{u}$, $\tilde{v}_{\bm{p}}(t)$ are defined in the main text (see Eq. (\ref{Eq_Coefficients})).
The dissipation term is $\Gamma_{\bm{p}\bm{p}^{\prime}}(t) = \sum_{n\in\mathbb{Z}}\Gamma^{nn}_{\bm{p}\bm{p}^{\prime}}(t)\exp[i(E_{\bm{p}}+n\Omega)t]$.

\end{widetext}

\bibliography{manuscript}

\end{document}